\documentclass[11pt]{article}
\usepackage{graphicx}
\usepackage[final]{acl}

\usepackage{times}
\usepackage{latexsym}

\usepackage[T1]{fontenc}
\usepackage[utf8]{inputenc}

\usepackage{microtype}

\usepackage{inconsolata}

\usepackage{graphicx}

\usepackage{amsmath}
\usepackage{amsfonts}
\usepackage{booktabs}

\usepackage{ragged2e}
\usepackage{listings}
\usepackage{tcolorbox}
\tcbset{
  colback=gray!5,
  colframe=black!20,
  arc=4pt,
  boxrule=0.5pt,
  left=6pt,
  right=6pt,
  top=6pt,
  bottom=6pt
}

\usepackage{tikz}
\usepackage{pgfplots}
\pgfplotsset{compat=1.18}

\title{ClusterRAG: Cluster-Based Collaborative Filtering for  Personalized Retrieval-Augmented Generation}
\author{
 \textbf{Gibson Nkhata\textsuperscript{1}},
 \textbf{Uttamasha Anjally Oyshi \textsuperscript{1}},
 \textbf{Quan Mai\textsuperscript{2}},
  \textbf{Susan Gauch\textsuperscript{1}}
\\
  \textsuperscript{1}University of Arkansas, Fayetteville, AR 72701, USA\\
  \textsuperscript{2}Walmart Inc., Bentonville, AR 72716, USA
\\
  \small{
    \{gnkhata,\,  uoyshi,\, sgauch\}@uark.edu
  }\\
  \small{quan.mai@walmart.com}
}

\begin{document}
\maketitle
\begin{abstract}

Personalized Retrieval-Augmented Generation (RAG) relies on accurately selecting user-relevant documents. In practice, existing RAG approaches often suffer from high retrieval costs and overlook that collaborative signals from similar users can enhance personalized generation for the current user. We propose \textbf{ClusterRAG}, a \textbf{Cluster}-Based Collaborative Filtering for Personalized \textbf{R}etrieval-\textbf{A}ugmented \textbf{G}eneration. ClusterRAG represents users through their profile documents, organizes users into semantically coherent clusters using density-based clustering, and performs retrieval at both the cluster and document levels via cluster-level similarity and fine-grained ranking. Extensive experiments on the LaMP benchmark demonstrate that jointly leveraging the target user’s profile and profiles from top similar users consistently yields the best performance across diverse tasks. Further analysis shows that ClusterRAG integrates seamlessly with different dense retrievers and rankers, and remains effective when paired with both fine-tuned and zero-shot language models.
\end{abstract}

\section{Introduction}

Retrieval-Augmented Generation (RAG) has emerged as a powerful paradigm for knowledge-intensive language tasks by combining parametric knowledge in large language models (LLMs) with non-parametric retrieval over external documents, significantly reducing hallucinations and improving factuality in text generation~\citep{lewis2020retrieval}. RAG systems typically retrieve documents related to the immediate query, then condition a generative model on those documents to produce responses~\citep{10.1145/3637528.3671470, Huang2024ASO, LI2025100417}. Despite impressive gains, current RAG pipelines often ignore long-term user information and inter-user relationships when constructing retrieval contexts, limiting personalization and the ability to leverage analogous users’ knowledge for improved generation quality.

Personalization is crucial in many real-world applications (e.g., personal assistants, tutoring systems, and personalized search) because user history and preferences strongly influence what information is relevant and how it should be framed~\citep{li2025surveypersonalizationragagent, ahmadcolbert}. Existing personalization strategies in RAG largely fall into two extremes: (1) user-only approaches that condition retrieval and prompts solely on a user’s own profile, which can be sparse or noisy~\citep{salemi-etal-2024-lamp,zerhoudi2024personaragenhancingretrievalaugmentedgeneration, 10.1145/3696410.3714717}, and (2) non-personalized approaches that ignore user history altogether~\cite{lewis2020retrieval, asai2024selfrag, yang2025heteragheterogeneousretrievalaugmentedgeneration,zhang2025rag}. Both approaches miss a middle ground where signals from similar users can enrich prompts while preserving user-specific nuance. Recent surveys~\cite{xu-etal-2025-personalized, li2025surveypersonalizationragagent} highlight the potential of end-to-end personalization across the RAG pipeline, from pre-retrieval user modeling to retrieval and generation, but also emphasize practical challenges such as balancing personal vs. collaborative signals.

Collaborative filtering (CF), long established in recommender systems~\cite{ijcai2017p447, 10.1145/3331184.3331267, liu2025user_similarity_review,  10.1145/3726302.3730075}, naturally complements personalization by exploiting similarities between users to infer missing preferences or relevant content. However, direct application of CF to RAG introduces new questions: (1) \textit{how should users be represented for retrieval tasks?}
(2) \textit{how to retrieve similar users 
to capture heterogeneous behavior at scale?} and (3) \textit{how to leverage both collaborative documents and  target user’s profile when forming prompts for LLMs? }

In this paper, we propose \textbf{Cluster}-Based Collaborative Filtering for  Personalized \textbf{R}etrieval-\textbf{A}ugmented \textbf{G}eneration (\textbf{ClusterRAG}), a practical pipeline that (1) constructs compact user representations by aggregating each user’s profile documents into embeddings, (2) groups users into clusters using  HDBSCAN~\citep{mcinnes2017hdbscan} to reveal cohorts of similar users and builds a cluster-level ranking matrix by scoring intra-cluster user similarities with effective rankers, e.g., ColBERT~\citep{10.1145/3397271.3401075}, and (3) clusters and retrieves candidate profile documents from the top $k$ similar users to form collaborative, user-only, or hybrid prompts for downstream generation. Our method explicitly leverages cluster structure to reduce search complexity, provide robust neighbor selection in variable-density settings, and enable principled mixing of collaborative and individual signals when constructing prompts. We evaluate multiple retrievers (ColBERT, Contriever~\cite{Izacard2021UnsupervisedDI}, BGE~\citep{10.1145/3626772.3657878}, BM25~\citep{inproceedings}, Recency and Random). Beyond methodological novelty, we also demonstrate that ClusterRAG is model-agnostic and robust across architectures and retrieval backbones. ClusterRAG integrates seamlessly with both fine-tuned sequence-to-sequence (seq2seq) encoder-decoder language models and zero-shot LLMs, without requiring any model-specific adaptation.

 We validate ClusterRAG on the LaMP benchmark~\citep{salemi-etal-2024-lamp} and report improvements in personalized generation quality compared to non-personalized and naive user-only baselines.  We also provide a link to an anonymous project repository for ClusterRAG
 \footnote{\url{https://github.com/academicprojects44/anonymous}}.

The remainder of this paper unfolds as follows. The next section presents related work; Section~\ref{sec:problem_formulation} provides problem formulation; Section~\ref{sec:model_framework} describes the ClusterRAG framework; Section~\ref{sec:experiments} presents the experimental evaluation; and Section~\ref{conclusion} presents the conclusion, limitations, and ethical considerations. Additional analysis and results are provided in the Appendix.

\section{Related Work}\label{se:related_work}
\textbf{Retrieval-Augmented Generation (RAG).}
The adoption of RAG has demonstrated improvements across a range of tasks, including question answering, dialogue comprehension, and code generation~\citep{lewis2020retrieval, Xu2023RECOMPIR, zerhoudi2024personaragenhancingretrievalaugmentedgeneration, 10.1145/3637528.3671470}.  Early RAG pipelines typically index a shared corpus (e.g., Wikipedia or domain corpora) and retrieve passages conditioned only on the current query; the retrieved passages are then used to condition an  LLM at generation time~\citep{lewis2020retrieval, zhang2025rag}. Subsequent work has explored numerous improvements to retriever architectures, reranking strategies, and retrieval-generation coupling mechanisms~\citep{Gao2023RetrievalAugmentedGF, siriwardhana-etal-2023-improving, 10.1145/3637528.3671470}. Even though these works establish strong task-agnostic baselines and sophisticated retriever-generator interfaces, they typically operate in a \textit{user-agnostic} manner and do not leverage a user’s long-term profile or cross-user signals when selecting documents for conditioning.

\textbf{Personalized Retrieval-Augmented Generation.}
A growing body of work studies how RAG can be adapted for personalized applications (e.g., personal assistants, tutoring, and individualized question answering) by incorporating user history, preferences, or authoring signals into retrieval and prompting~\citep{salemi-etal-2024-lamp, zerhoudi2024personaragenhancingretrievalaugmentedgeneration, li2025surveypersonalizationragagent}. Some systems personalize LLMs by (1) fine-tuning model parameters (either fully or selectively) for individual users~\citep{li-liang-2021-prefix, hu2022lora, zollo2025personalllm}, (2) incorporating latent user representations into the model~\cite{10.1145/3701716.3715463, qiu-etal-2025-latent, huber2025embeddingtoprefixparameterefficientpersonalizationpretrained}, and (3) augmenting model prompts with a user profile or a small set of user documents~\citep{10.1145/3477495.3531722, salemi-etal-2024-lamp, 10.1145/3626772.3657783}. The first two strategies require modifying the model’s architecture or parameters, which can be expensive, or in some cases infeasible, due to storage, computational, and time constraints. In addition, they cannot perform well for cold-start users.  In contrast, the third approach, which is adopted in this work, can be applied to any generative model~\cite{10.1145/3626772.3657783}. User-specific personalization works well when profiles are dense and representative; this notwithstanding, user-only personalization suffers when profiles are sparse, noisy, or unrepresentative of the current intent.

\textbf{Collaborative Personalized Retrieval Augmented Generation.}
Collaborative filtering has been extensively studied in recommender systems and consistently shown to be effective~\citep{ijcai2017p447, 10.1145/3331184.3331267, Zhang2024QAGCFGC, Shen2024ASO, 10.1145/3626772.3657811, Tang2025ThinkBR, 10.1145/3715099, Zhang2025TestTimeAF}. The core premise is that users with comparable interaction histories tend to exhibit similar preferences; therefore, leveraging items favored by similar users can help generate relevant recommendations for a target user. Recent efforts have started to marry CF and retrieval for generation~\citep{10.1145/3726302.3730075, 10.1145/3696410.3714908}. For instance, \citet{10.1145/3726302.3730075} employs contrastive learning to generate user embeddings that retrieve similar users and incorporate collaborative signals with a user input for prompt creation. 

Despite this progress, two practical challenges remain unresolved in current collaborative RAG research. First, naively computing pairwise similarities across millions of users is costly; clustering users into cohorts can reduce search complexity. ClusterRAG is designed to address this issue by combining user-level clustering with document-level collaborative retrieval and introducing cluster-level ranking matrices that summarize intra-cluster user similarity and thereby enabling robust neighbor selection even in variable-density cohorts. Second, once neighbor users are found, selecting which of their documents to include and how to merge collaborative documents with a target user’s own profile remains an open design choice with direct impact on generation quality. ClusterRAG uses flexible prompt fusion modes and evaluates multiple retrievers, showing empirical robustness across both fine-tuned and training-free generative models.  

To the best of our knowledge, ClusterRAG is the first framework to integrate user-level clustering with collaborative document retrieval for personalized RAG, explicitly leveraging cross-user similarity to enrich sparse user profiles for personalized generation.

\section{Problem Formulation}\label{sec:problem_formulation}

A standard RAG setup consists of two core components, retrieval and generation: given an input query $x$, the model predicts the most probable output sequence $y$ conditioned on $x$ and a retrieved document $d$. Personalized RAG extends this formulation by conditioning generation on a user $u$, typically represented through a user profile. Formally, we let 
$T = \{(u_1, x_1, y_1), (u_2, x_2, y_2), \ldots, (u_N, x_N, y_N)\}$ denote a set of $N$ training instances, where each tuple consists of a user $u$, a user-issued input query $x$, and a corresponding personalized ground-truth output $y$. For each user $u$, a user profile $U_p$ is available and serves as an auxiliary context for personalized generation. The profile $U_p = \{d_1, d_2, \ldots, d_n\}$ is a collection of personal documents or historical records associated with $u$, such as past queries and generated outputs.

In ClusterRAG, given a target user $u$, our objective is to identify the top $k$ most similar users using a clustering-assisted ranking strategy. We then retrieve and rank the top $m$ documents from these users using a retriever (ranker) $R$. 

\section{ClusterRAG Framework}\label{sec:model_framework}

\begin{figure*}[t]
  \includegraphics[width=1.0\linewidth]{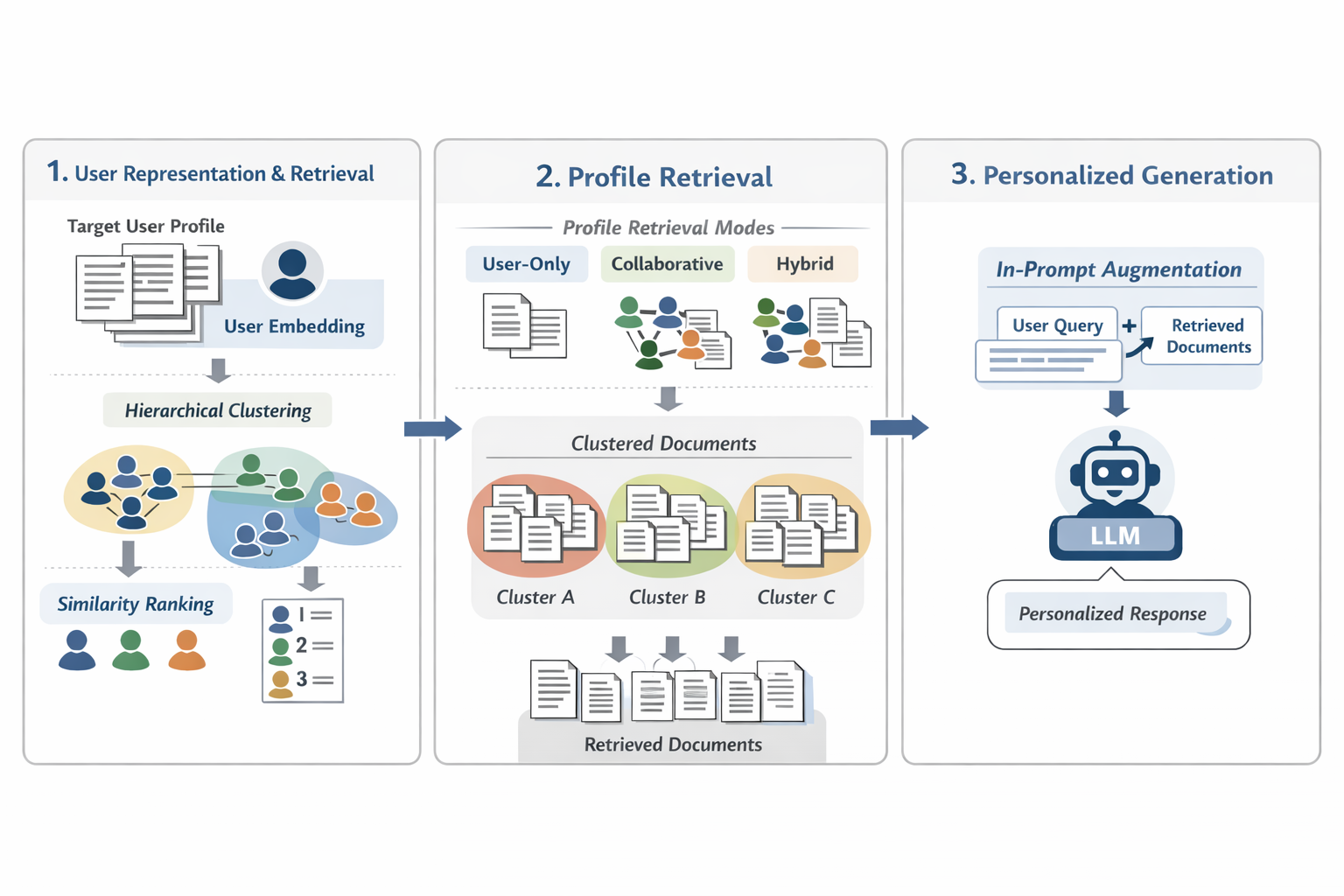}
  \caption{Overview of the ClusterRAG framework.}
  \label{fig:model_framework}
\end{figure*}

ClusterRAG is built on the intuition that users with similar behaviors and preferences can provide valuable contextual signals for one another. By combining information from a target user’s own profile with carefully selected profiles from similar users, ClusterRAG enhances an LLM’s ability to generate accurate and personalized responses. As shown in Figure~\ref{fig:model_framework}, ClusterRAG framework consists of three main stages: (1) user representation and retrieval, (2) profile retrieval, and (3) personalized generation.

\subsection{User Representation and Retrieval}\label{subsec:user_retrieval}

Since explicit user representations are typically unavailable, we first construct user embeddings from observed interaction data. All data instances associated with a user are aggregated into a user-level profile $U_p$. Each document $d_i \in U_p$ is encoded using a dense embedding model, specifically ColBERTv2~\citep{santhanam-etal-2022-colbertv2}. A compact user representation is then obtained by averaging document embeddings:

\begin{equation}\label{eq:user_emb_rep}
    \mathbf{z}_u = \frac{1}{n_u} \sum_{i=1}^{n_u} f(d_i),
\end{equation}
where $f(\cdot)$ denotes the embedding function. % and $\mathbf{z}_u$ is the resulting user embedding.

Computing similarities between all user pairs is prohibitively expensive at scale. To address this, ClusterRAG groups users into similarity-based cohorts using a hierarchical density-based clustering method,   HDBSCAN~\citep{mcinnes2017hdbscan}, which automatically identifies clusters of varying density, making it well-suited for collaborative filtering in our setting, where the number of user groups in the user-document collection is unknown a priori. 

Clustering alone does not quantify the relative similarity of users within each cluster. Therefore, we compute intra-cluster similarities using a modern reranker, ColBERTv2~\citep{santhanam-etal-2022-colbertv2}, which provides fine-grained token-level interactions inherited from ColBERT~\citep{10.1145/3397271.3401075} while incorporating residual compression and denoised supervision for improved efficiency and generalization. These properties make ColBERTv2 well-suited for robust similarity estimation between user profiles.

\paragraph{Cluster-Level Similarity Ranking.}
This step aims at restricting similarity computation to cluster members to improve robustness and scalability by focusing comparisons on behaviorally consistent cohorts. For each cluster $C$, we construct an intra-cluster similarity matrix $R^C$ defined as:
\begin{equation}\label{eq:inter_cluster_sim}
    R^{C}_{u,v} = ColBERTv2(\mathbf{z}_u, \mathbf{z}_v),
\end{equation}
where $u,v \in C$ and $v \neq u$. The diagonal entries in $R^C$ correspond to self-similarity and are discarded. For each user $u$, we finally retain an ordered list of the top $k$ most similar users within the same cluster. 

\subsection{Profile Retrieval}\label{subsec:profile_retrieval}

The profile retrieval stage integrates search and ranking to identify documents that are most beneficial for personalized generation~\citep{Huang2024ASO}. Incorporating collaborative filtering introduces two key challenges: (1) selecting relevant documents from similar users and (2) effectively leveraging both collaborative documents and the target user’s documents for personalized RAG. ClusterRAG addresses these challenges by leveraging cluster structure to provide both topical coherence and retrieval efficiency and organizes profile retrieval as follows.

\paragraph{Profile Retrieval Modes.}
First, given a query $q$ from user $u$, we retrieve candidate profile documents using one of the following three retrieval modes.
(1) \emph{User-only retrieval}: the simplest strategy, which considers only the user’s own profile.
(2) \emph{Collaborative retrieval}: this mode retrieves documents from the profiles of the top $k$ most similar users. It is particularly beneficial for sparse or cold-start users, whose own profiles may not adequately capture the intent of the current query.
(3) \emph{Hybrid retrieval}: this mode combines both user-only and collaborative profiles. As a result, the effective profile $U_p$ used for generation may consist of: (i) documents from the target user only, (ii) documents from similar users only,  or (iii) a combination of documents from both sources.

\paragraph{Clustering for Topical Organization.}
After selecting a profile retrieval mode,  all candidate profile documents are encoded using a dense retriever, such as ColBERTv2, % Contriever~\citep{Izacard2021UnsupervisedDI}, or BAAI General Embeddings (BGE)~\citep{10.1145/3626772.3657878}, 
and partitioned into clusters using HDBSCAN, producing a set of clusters $\mathcal{C} = \{C_1, \ldots, C_K\}$ with corresponding computed centroids $\mathcal{M} = \{\boldsymbol{\mu}_1, \ldots, \boldsymbol{\mu}_K\}$. This clustering captures latent topical structure and enables ClusterRAG to automatically infer the number of topics present in a user’s profile without requiring prior specification.

\paragraph{Cluster-Level Indexing and Retrieval.}
Finally, ClusterRAG employs a two-stage retrieval strategy. First, a cluster index stores centroid embeddings: given a query $q$, its embedding $e_q$ is computed similarly as document embeddings,  and compared against all centroids, and the top $B$ clusters are selected from $\mathcal{C}$. Second, within each selected cluster, documents are retrieved and reranked by similarity to $e_q$, and the top $m$ documents are selected for generation.

This hierarchical retrieval reduces complexity from $\mathcal{O}(N)$ to $\mathcal{O}\!\left(K + B \cdot N / K\right)$, where $N$ is the total number of documents, each cluster $C_i$ contains $\approx|N/K|$ documents, and $B \leq K \leq N$.

We use ColBERTv2 as the primary model to retrieve and rerank profile documents; even so, our experiments evaluate additional dense, sparse, heuristic, and random retrievers to demonstrate the framework’s retriever-agnostic design.

In cold-start scenarios, where no historical user documents are available within the system, the user embedding for similarity computation is derived directly from the current query. Under typical conditions, user similarity is estimated using embeddings aggregated from the user’s available profile documents. When user history remains sparse, ClusterRAG inherently defaults to either user-only retrieval or a hybrid retrieval strategy, thereby maintaining robust performance despite limited contextual information.

\subsection{Personalized Generation}
Effective generation by LLMs critically depends on well-engineered prompts that are tailored to the downstream task. Prompts allow seamless integration of pre-trained models into downstream tasks by eliciting desired model behaviors solely based on the given prompt~\cite{sahoo2025systematicsurveypromptengineering}. To effectively leverage selected user documents from $U_p$, ClusterRAG adopts \emph{In-Prompt Augmentation} \textit{(IPA)}~\cite{salemi-etal-2024-lamp}, which integrates the user query with all relevant retrieved documents directly within the prompt. IPA is particularly well suited for ClusterRAG, as it can be applied to both training-free (zero-shot) and fine-tuned settings and is compatible with a wide range of model architectures. Accordingly, ClusterRAG supports both fine-tuned LLMs and zero-shot generative models and can be combined with a variety of state-of-the-art document retrievers.

Specifically, to balance user profile context and query specificity,
given a maximum prompt length $L_{\max}$, the allocated profile length is computed as:
\begin{equation}\label{eq:prompt_len}
    |U_p| = \mathcal{G}_{t}\!\left(L_{\max} - \min\!\left(|q|, \lfloor \gamma L_{\max} \rfloor\right)\right),
\end{equation}
where $\gamma \in [0,1]$ is a tunable mixing parameter, $|q|$ is query length, and $\mathcal{G}_{t}(\cdot)$ is a task-specific prompt generator (we provide detailed task-specific prompt generators in Section~\ref{subsec:prompts_main_sec} and Appendix~\ref{sec:prompts_appendix}). This formulation allows ClusterRAG to incorporate strong personalization signals while preserving the relevance of the current query.

\section{Experiments}\label{sec:experiments}
In this section, we present the experimental setup employed with ClusterRAG, including the datasets, baseline models, evaluation metrics, implementation details, and experimental results. We also provide an overview of the prompts used in our experiments.

\subsection{Experimental Setup}\label{exp_setup}
\paragraph{Datasets.}
Our experiments employ the LaMP benchmark~\citep{salemi-etal-2024-lamp}, a publicly available dataset that covers a broad range of personalized text generation tasks. The benchmark consists of three personalized text classification tasks and four personalized text generation tasks. One of the four text generation tasks, \textbf{LaMP-6} ( Personalized Email Subject Generation), is excluded in this work because its data is not publicly available. Specifically, the remaining tasks include: \textbf{LaMP-1:} \textit{Personalized Citation Identification}, formulated as a binary classification task; \textbf{LaMP-2:} \textit{Personalized Movie Tagging}, a 15-class categorical classification task; \textbf{LaMP-3:} \textit{Personalized Product Rating}, an ordinal classification task predicting ratings from one to five stars for e-commerce products; \textbf{LaMP-4:} \textit{Personalized News Headline Generation}; \textbf{LaMP-5:} \textit{Personalized Scholarly Title Generation};  and \textbf{LaMP-7:} \textit{Personalized Tweet Paraphrasing}.  ClusterRAG experiments follow the time-based LaMP split to partition the data into training, validation, and test sets. We provide statistics of the dataset in Table~\ref{tab:lamp_stats}, detailed dataset statistics in Table~\ref{tab:lamp_stats_appendix} in Appendix~\ref{sec:detailed_lamps_statistics}, and detailed task descriptions in Appendix~\ref{app_sec:appendix1}. We additionally provide dataset licensing information in Appendix~\ref{sec:detailed_lamps_statistics}.

\begin{table}[t]
\centering
\begin{tabular}{lcccc}
\toprule
\textbf{Task} & \textbf{\#users} & \textbf{\#train} & \textbf{\#dev} & \textbf{\#test} \\
\midrule
LaMP-1 & 6542  & 6542  & 1500 & 1500 \\
LaMP-2 & 929   & 5073  & 1410 & 1557 \\
LaMP-3 & 20000 & 20000 & 2500 & 2500 \\
LaMP-4 & 1643  & 12500 & 1500 & 1800 \\
LaMP-5 & 14682 & 14682 & 1500 & 1500 \\
LaMP-7 & 13437 & 13437 & 1498 & 1500 \\
\bottomrule
\end{tabular}

\caption{Statistics of the LaMP benchmark with time-based data split.}
\label{tab:lamp_stats}
\end{table}

\paragraph{Evaluation Metrics.} Following previous work~\citep{salemi-etal-2024-lamp, 10.1145/3626772.3657783, 10.1145/3726302.3730075}, we evaluate LaMP-1 and  LaMP-2 using Accuracy and F1-measure, and LaMP-3 using Mean Absolute Error (MAE) and Root Mean Squared Error (RMSE). We evaluate text generation performance on LaMP-4, LaMP-5, and LaMP-7 using ROUGE-1 (R-1) and ROUGE-L (R-L)~\citep{lin-2004-rouge}. 

\paragraph{Baseline Models.} We firstly compare ClusterRAG to \textbf{no personalization} and  call this \textbf{vanillaRAG}. In this baseline, the generative model is presented with the original task’s input without any profile documents to assess whether personalization improves the model effectiveness. 
Then we consider personalized baselines: (1) \textbf{User-only} models, which include (a) \textbf{ROPG}~\citep{10.1145/3626772.3657783}: it optimizes the dense retrieval model based on the results generated by a LLM and (b) \textbf{LaMP-IPA}: it was introduced with the LaMP benchmark, and it uses in-prompt augmentation for prompt creation; 
(2) \textbf{Collaborative} baseline,  \textbf{CFRAG}~\citep{10.1145/3726302.3730075}, that uses contrastive learning to find similar users. To the best of our knowledge, CFRAG is the only existing collaborative work for personalized RAG; therefore, we compare these baselines with three versions of ClusterRAG: user-only, collaborative, and hybrid profile retrieval modes. 

\paragraph{Implementation details.}
We implement ClusterRAG using the HuggingFace transformers framework~\citep{wolf-etal-2020-transformers} and the PyTorch library ~\citep{paszke2019pytorch}. ClusterRAG adopts a finetuned  FlanT5-base~\citep{JMLR:v25:23-0870} for generation; unless explicitly
stated otherwise (in experiments with zero-shot LLMs), it uses a causal Qwen2-7B-Instruct~\citep{yang2024qwen2technicalreport} or a seq2seq FlanT5-XXL~\citep{JMLR:v25:23-0870}; all models are open-source. Training is performed using the Trainer or Seq2SeqTrainer APIs\footnote{\url{https://github.com/huggingface/transformers}}, depending on whether the LLM backbone is a causal or seq2seq model. FlanT5-base has 250M parameters; Qwen2-7B-Instruct uses 7.07B parameters; and FlanT5-XXL has 11B parameters.

We use AdamW~\cite{loshchilov2019decoupledweightdecayregularization} optimization with a learning rate of $5\times10^{-5}$, weight decay of $10^{-4}$, linear learning rate scheduling, and a warm-up ratio of $0.05$. Models are trained for up to 30 epochs, with evaluation and checkpointing conducted at the end of each epoch. The maximum prompt and output lengths ($L_{\max}$ and $|\bar{y}|$) of generative models is set to 512 and 128 tokens, respectively. Maximum sequence length for embedding models is set to 256 tokens. We set the number of collaborative (similar) users, $k$, to 1 and retrieved profile documents, $m$,  to 2. 
 $\gamma$ for prompt formulation in Equation~\ref{eq:prompt_len} is set to 0.55, while batch size is set to 16. Beam search~\cite{freitag-al-onaizan-2017-beam}  with a beam size of 4 is employed for text generation. All hyperparameters for the baseline models are searched according to the settings in the original papers. We select the optimal hyperparameters for ClusterRAG via grid search and report the corresponding tuning grid in Table~\ref{tab:hyperparam_tune_grid} in Appendix~\ref{sec:appendix2}.  All experiments are conducted on a Quadro RTX 8000 GPUs, 48 GB VRAM for a range of 10-24 hours per experiment depending on the task. 
 
\subsection{Prompts Used in ClusterRAG}\label{subsec:prompts_main_sec}
This subsection presents the prompt templates employed during generation. Each prompt contains an instruction, input, and profile. We provide the template for LaMP-1 only below; templates for other tasks are presented in Appendix~\ref{sec:prompts_appendix}. In the template, \{\textit{Paper abstract}\} and \{\textit{Reference list}\}, \{\textit{Movie description}\}  represent user input for the corresponding LaMP tasks, while \{\textit{Paper abstract}\} represent user profile entries. The remaining text is the instruction guiding an LLM to generate the intended output.

\begin{tcolorbox}
\textbf{LaMP-1 Prompt Template:} Given an author  who has previously written papers \{\textit{Paper list}\}  
and now has written \{\textit{Paper abstract}\}. Which reference below is related? Just answer with [1] or [2] without explanation. \{\textit{Reference list}\}
\end{tcolorbox}

\subsection{Experimental Results}\label{subsec:exp_results}
When reporting experimental results, we identify statistically significant differences of ClusterRAG performance using a two-tailed paired t-test for generation and ordinal classification evaluation (ROUGE-1, ROUGE-L, MAE, and RMSE) and McNemar test for categorical text classification evaluation (Accuracy and F1).  We report results comparing ClusterRAG with baselines, analyzing its retriever-agnostic design, language model versatility, and ablation study. In all tables, the symbol $\uparrow$ indicates that higher values are better, while the symbol $\downarrow$ indicates that lower values are better; all results presented are obtained from a single experimental run.

\subsubsection{Comparison with Baselines} Comparison results are presented in Table \ref{tab:baseline_results}, which shows that ClusterRAG consistently outperforms all baselines across the LaMP benchmark. Importantly, the hybrid variant (\textit{ClusterRAG-H}) achieves the best performance on every task. These improvements indicate that retrieving and aggregating documents from similar users substantially enhances RAG, providing more relevant and personalized evidence than standard RAG pipelines. Notably, ClusterRAG achieves strong performance using only two profile documents, whereas baseline methods require at least four documents to reach their optimal results, indicating that ClusterRAG is well suited for low-resource personalization settings. While \textit{ClusterRAG-C }(collaborative) and \textit{ClusterRAG-U} (user-only) achieve competitive second-best results on several tasks, their combination in the hybrid model yields the most robust and consistent gains across diverse task settings. 

\begin{table*}[t]
\centering
\setlength{\tabcolsep}{3.8pt}
\renewcommand{\arraystretch}{1.15}
\footnotesize
\begin{tabular}{lcccccccccccc}
\toprule
\textbf{Models} 
& \multicolumn{2}{c}{\textbf{LaMP-1}} 
& \multicolumn{2}{c}{\textbf{LaMP-2}} 
& \multicolumn{2}{c}{\textbf{LaMP-3}} 
& \multicolumn{2}{c}{\textbf{LaMP-4}} 
& \multicolumn{2}{c}{\textbf{LaMP-5}} 
& \multicolumn{2}{c}{\textbf{LaMP-7}} \\
\cmidrule(lr){2-3}\cmidrule(lr){4-5}\cmidrule(lr){6-7}
\cmidrule(lr){8-9}\cmidrule(lr){10-11}\cmidrule(lr){12-13}

& Acc.$\uparrow$ & F1$\uparrow$
& Acc.$\uparrow$ & F1$\uparrow$
& MAE$\downarrow$ & RMSE$\downarrow$
& R-1$\uparrow$ & R-L$\uparrow$
& R-1$\uparrow$ & R-L$\uparrow$
& R-1$\uparrow$ & R-L$\uparrow$ \\
\midrule
vanillaRAG &0.630  &0.630  &0.520  &0.440  &0.371 &0.709  &0.171  &0.154  &0.462  &0.413  &0.310  &0.273  \\
LaMP-IPA & \underline{0.674} & \underline{0.664} & \underline{0.570} & \underline{0.522} &\underline{0.289}  &\underline{0.608}  &0.175  &0.169  &0.472  &0.423  &\underline{0.508}  &\underline{0.457}  \\
ROPG             & 0.644 & 0.322 & 0.468 & 0.031 & 0.346 & 0.692 &\underline{0.184} & \underline{0.163} & 0.464 & 0.396 & 0.353 & 0.288 \\
CFRAG  & 0.633 & 0.327 & 0.534 & 0.036 & 0.354 & 0.707 & 0.162 & 0.141 & \underline{0.473} & \underline{0.425} & 0.375 & 0.306 \\

ClusterRAG-C  &0.674$^{*}$  &0.673$^{*}$  &0.644  &0.607  &0.284  &0.624  &0.179  &0.157 &0.480$^{*}$  &0.430$^{*}$  &0.507  &0.454  \\

ClusterRAG-U       &0.645  &0.645 &0.649$^{*}$ &0.612$^{*}$ &0.271$^{*}$ &0.599$^{*}$ &0.184$^{*}$  &0.165$^{*}$ &0.475  &0.425 &0.514$^{*}$  &0.464$^{*}$ \\
\textbf{ClusterRAG-H}
                 & \textbf{0.690} & \textbf{0.690} 
                 & \textbf{0.661} & \textbf{0.620} 
                 & \textbf{0.270} & \textbf{0.594} 
                 & \textbf{0.190} & \textbf{0.176} 
                 & \textbf{0.490} & \textbf{0.440} 
                 & \textbf{0.521} & \textbf{0.470} \\
\bottomrule
\end{tabular}
\caption{Comparison of the performance of ClusterRAG with baselines on the LaMP benchmark.  Best results are shown in \textbf{bold} and second-best in \underline{underlined}; boldface indicates statistically significant improvements over the second-best ($p < 0.05$). The symbol $^{*}$ denotes the second-best ClusterRAG variant.}
\label{tab:baseline_results}
\end{table*}

\subsubsection{ClusterRAG Retriever-Agnostic Design}
\begin{figure}[t]
    \centering
    \footnotesize
    \begin{tikzpicture}
        \begin{axis}[
            ybar,
            bar width=8pt,
            width=0.8\columnwidth,
            height=0.4\columnwidth,
            ymin=0.46, ymax=0.50,
            ylabel={ROUGE-1},
            symbolic x coords={
                Ran,
                Rec,
                BM25,
                Con,
                BGE,
                Col
            },
            xtick=data,
            xticklabel style={rotate=30, anchor=east},
            nodes near coords,
            nodes near coords style={font=\scriptsize},
            enlarge x limits=0.15,
            axis lines=left,
            tick align=outside,
            grid=both,
            grid style={dashed, gray!30},
        ]
        \addplot coordinates {
            (Ran,0.470)
            (Rec,0.481)
            (BM25,0.483)
            (Con,0.480)
            (BGE,0.482)
            (Col,0.490)
        };
        \end{axis}
    \end{tikzpicture}
    \caption{Retrivers' ROGUE-1 scores on LaMP-5.}
    \label{fig:lamp5_r1_hist}
\end{figure}
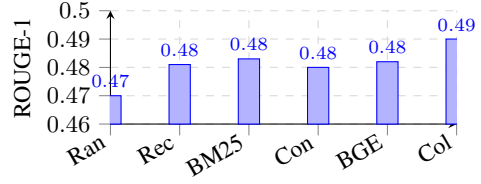

In addition to \textit{ColBERTv2}, ClusterRAG explores five more retrievers: (1) a dense unsupervised dual-encoder retriever, \textit{Contriever}~\citep{Izacard2021UnsupervisedDI}, (2) a fine-tuned multilingual dense retriever optimized for semantic similarity and retrieval tasks, \textit{BGE}~\citep{10.1145/3626772.3657878}, (3) a classical sparse lexical retriever based on term frequency-inverse document frequency (TF-IDF), \textit{BM25}~\citep{inproceedings}, (4) a heuristic retriever that ranks documents solely based on temporal proximity to the query time, favoring the most recently published documents, \textit{Recency}, and (5)
a non-informative baseline that samples documents uniformly at random, \textit{Random}. We provide retriever-agnostic design results in Table~\ref{tab:retriever_agnostic_results} for LaMP-(1,2,7) and Figure~\ref{fig:lamp5_r1_hist} for LaMP-5. 
The table and figure demonstrate that ClusterRAG consistently benefits from stronger retrievers, with dense semantic models outperforming sparse and heuristic baselines across all LaMP tasks. \textit{ ColBERTv2 (ColBERT or Col)} achieves the best overall performance on all tasks, highlighting the advantage of late-interaction matching in personalized retrieval. \textit{BGE} and \textit{Contriever (Con)} provide competitive performance, confirming the retriever-agnostic nature of ClusterRAG, while \textit{BM25} and \textit{Recency (Rec)} offer modest gains over \textit{Random (Ran)} but lag behind dense methods. These results indicate that ClusterRAG is robust across retrieval paradigms, yet most effectively leverages high-capacity dense retrievers to maximize personalization and generation quality.

\begin{table}[t]
\centering
\setlength{\tabcolsep}{3.8pt}
\renewcommand{\arraystretch}{1.15}
\footnotesize
\begin{tabular}{lcccccc}
\toprule
\textbf{Retrievers} 
& \multicolumn{2}{c}{\textbf{LaMP-1}} 
& \multicolumn{2}{c}{\textbf{LaMP-2}} 
& \multicolumn{2}{c}{\textbf{LaMP-7}} \\
\cmidrule(lr){2-3}\cmidrule(lr){4-5}\cmidrule(lr){6-7}

& Acc.$\uparrow$ & F1$\uparrow$
& Acc.$\uparrow$ & F1$\uparrow$
& R-1$\uparrow$ & R-L$\uparrow$ \\
\midrule
Random 
& 0.640 & 0.639 
& 0.609 & 0.608 
& 0.500 & 0.449 \\

Recency 
& 0.659 & 0.650 
& 0.618 & 0.610 
& 0.507 & 0.456 \\

BM25 
& 0.662 & 0.658 
& 0.629 & 0.621 
& 0.510 & 0.460 \\

Contriever  
& 0.681 & 0.681 
& 0.649 & \underline{0.623} 
& \underline{0.511} & 0.459 \\

BGE  
& \underline{0.684} & \underline{0.682} 
& \underline{0.658} & 0.613 
& 0.509 & \underline{0.461} \\

\textbf{ColBERT}
& \textbf{0.690} & \textbf{0.690} 
& \textbf{0.661} & \textbf{0.620} 
& \textbf{0.521} & \textbf{0.470} \\
\bottomrule
\end{tabular}
\caption{Comparison of the performance of ClusterRAG under different retrievers on LaMP-(1,2,7).}
\label{tab:retriever_agnostic_results}
\end{table}

\subsubsection{ClusterRAG LLM Versatility}

Table~\ref{tab:llms} reports the performance of ClusterRAG on LaMP-(1,2,5) when paired with zero-shot LLMs: FlanT5-XXL and Qwen2-7B-Instruct. For each LLM, we compare a non-personalized variant (\textit{nFlan}, \textit{nQwen2}) against its personalized counterpart (\textit{pFlan}, \textit{pQwen2}).  As shown in Table~\ref{tab:llms}, personalized variants consistently outperform their non-personalized counterparts across all tasks, demonstrating the effectiveness of ClusterRAG in injecting user-specific and collaborative signals into generation. Notably, \textit{pFlan} achieves the strongest overall performance on LaMP-1,  while \textit{pQwen2} attains the best results on LaMP-2 and LaMP-5, indicating that the benefits of ClusterRAG generalize across model architectures, providing consistent gains without requiring additional model fine-tuning.

\begin{table}[t]
\centering
\setlength{\tabcolsep}{3.8pt}
\renewcommand{\arraystretch}{1.15}
\footnotesize
\begin{tabular}{lcccccc}
\toprule
\textbf{LLMs} 
& \multicolumn{2}{c}{\textbf{LaMP-1}} 
& \multicolumn{2}{c}{\textbf{LaMP-2}} 
& \multicolumn{2}{c}{\textbf{LaMP-5}} \\
\cmidrule(lr){2-3}\cmidrule(lr){4-5}\cmidrule(lr){6-7}

& Acc.$\uparrow$ & F1$\uparrow$
& Acc.$\uparrow$ & F1$\uparrow$
& R-1$\uparrow$ & R-L$\uparrow$ \\
\midrule
nFlan 
& 0.546 & 0.540 
& 0.451 & 0.448 
& 0.431 & 0.398 \\

pFlan 
& \textbf{0.648} & \textbf{0.647} 
& 0.601 & 0.601 
& 0.484 & 0.432 \\

nQwen2 
& 0.602 & 0.600 
& 0.521 & 0.521 
& 0.457 & 0.419 \\

pQwen2 
& 0.639 & 0.635 
& \textbf{0.610} & \textbf{0.606} 
& \textbf{0.488} & \textbf{0.447} \\
\bottomrule
\end{tabular}
\caption{Performance comparison of ClusterRAG using different LLMs on  LaMP-(1,2,5). }
\label{tab:llms}
\end{table}

\subsubsection{Ablation Study}
The integral components of ClusterRAG are user representation and retrieval and profile retrieval. By systematically removing or replacing these individual components, we assess their impact on personalized generation performance on selected LaMP tasks in Table~\ref{tab:ablation_study}.

First, we examine the role of collaborative user modeling. Replacing clustering-based neighbor selection with random user sampling (\textit{w/o user clustering}) leads to a substantial degradation across all tasks, highlighting the importance of structured user grouping. Similarly, removing intra-cluster similarity ranking (\textit{w/o intra-cluster sim}) consistently reduces performance, indicating that fine-grained similarity estimation within clusters is critical for identifying truly relevant collaborative signals. Lastly, we analyze the profile retrieval module. Using cluster centroids alone to represent user profiles (\textit{Centroids only}) results in the largest performance drop, demonstrating that document-level evidence is essential. Excluding document ranking (\textit{w/o doc ranking}) further degrades results, confirming that effective reranking is necessary to prioritize high-quality contextual evidence. Replacing HDBSCAN with  $k$\textit{-means}~\citep{5453745} clustering yields slightly weaker yet competitive performance, suggesting that while ClusterRAG is robust to the choice of clustering algorithm, density-aware clustering provides additional benefits.

\begin{table}[t]
\centering
\setlength{\tabcolsep}{3.8pt}
\renewcommand{\arraystretch}{1.15}
\footnotesize
\begin{tabular}{lcccc}
\toprule
\textbf{Derivatives} 
& \multicolumn{2}{c}{\textbf{LaMP-3}} 
& \multicolumn{2}{c}{\textbf{LaMP-7}} \\
\cmidrule(lr){2-3}\cmidrule(lr){4-5}

& MAE$\downarrow$ & RMSE$\downarrow$
& R-1$\uparrow$ & R-L$\uparrow$ \\
\midrule
w/o user clustering 
& 0.320 & 0.637 
& 0.458 & 0.371  \\

w/o intra-cluster sim 
& 0.329 & 0.639 
& 0.501 & 0.442  \\

w/o doc ranking  
& 0.331 & 0.642 
& 0.462 & 0.413  \\

Centroids only  
& 0.400 & 0.643 
& 0.472 & 0.438 \\

$k$-means 
& 0.291 & 0.610 
& 0.502 & 0.453  \\

\textbf{ClusterRAG}
& \textbf{0.270} & \textbf{0.594} 
& \textbf{0.521} & \textbf{0.470} \\
\bottomrule
\end{tabular}
\caption{Ablation study of ClusterRAG on LaMP-(3,7).}
\label{tab:ablation_study}
\end{table}

\subsection{Discussion}
\paragraph{Computational Effectiveness.}
ClusterRAG consistently improves personalized generation across diverse tasks and evaluation metrics by jointly leveraging user-specific history and collaborative signals from similar users. The gains observed in both classification and generation settings indicate that clustering-based collaborative filtering provides complementary information beyond individual user profiles, particularly for sparse or ambiguous queries. 

\paragraph{Computational Efficiency.}
ClusterRAG is computationally efficient due to its modular and lightweight design. User and document clustering is performed using HDBSCAN, a non-parametric algorithm without learnable parameters, enabling fast and scalable user grouping. The primary retriever, ColBERTv2, maintains a parameter size comparable to BERT by introducing only a small linear projection layer (approximately 0.1M parameters), resulting in a total model size of roughly 110M parameters. Since user and document embeddings can be precomputed offline, inference primarily involves similarity computations, yielding low latency and minimal overhead. This efficiency allows ClusterRAG to scale effectively and generalize rapidly to new datasets.

\paragraph{Collaborative User Feeddback.} As shown in our ablation study, the framework supports user-only retrieval and degrades gracefully when collaborative signals are limited. In cold-start or standalone assistant scenarios (e.g., systems such as OpenClaw~\citep{openclaw2025}), ClusterRAG can operate in a user-only mode without clustering. The collaborative component is modular and can be disabled or restricted depending on deployment constraints. 

\paragraph{Static User Embeddings and Evolving Profiles.}
The current implementation of ClusterRAG performs offline clustering for experimental clarity and reproducibility. However, ClusterRAG can be extended to dynamic settings through: (1) periodic batch re-clustering, (2) incremental clustering techniques, (3) online embedding updates with cluster reassignment, (4) maintaining cluster centroids and assigning new users without full recomputation. Since HDBSCAN supports soft clustering and incremental assignment strategies, the framework can adapt without recomputing all pairwise similarities.

\paragraph{Cluster Overlap and Boundary Users.} HDBSCAN naturally handles variable-density regions and noise points, boundary users can be assigned flexibly or treated as outliers. ClusterRAG assigns all outliers to a dedicated cluster.

\paragraph{Large Language Models Selection.} We used FlanT5-base, FlanT5-XXL, and Qwen2-7B-Instruct to demonstrate generalization across encoder-decoder and decoder-only architectures. Due to computational constraints, we focused on established open models to enforce reproducibility

%Supplemental experiments.
We further investigate the sensitivity of ClusterRAG to the number of similar users and the size of the retrieved profile  in Appendix~\ref{sec:appendx_impact_k_m}.
In addition, Appendix~\ref{sec_app:cluster_cohesion} evaluates cluster cohesion for collaborative user retrieval, and  Appendix~\ref{sec:case_study} presents a qualitative case study that illustrates the performance of ClusterRAG.%in producing personalized outputs.

\section{Conclusion}\label{conclusion}

This work introduced ClusterRAG, a collaborative framework that organizes users and their documents into semantically coherent clusters and performs retrieval at both the cluster and document levels, effectively reducing search complexity while preserving retrieval quality. Extensive experiments on the LaMP benchmark demonstrate that the hybrid profile retrieval mode, which jointly leverages the target user’s profile and profiles from top similar users, is the most effective configuration, yielding the best overall performance. Additionally, the experiments indicate that ClusterRAG is retriever-agnostic, allowing seamless integration with different dense retrievers and rankers, and remains effective when paired with both fine-tuned and zero-shot language models, highlighting its robustness and generality. Overall, ClusterRAG offers a design approach that improves effectiveness without incurring significant computational overhead.%, making it a practical and extensible solution for personalized RAG.

\section*{Limitations}\label{sec:limitations}
While ClusterRAG's primary goal is to retrieve similar-user documents and enhance personalized RAG, we highlight a few factors that may affect the model's performance.
First, ClusterRAG relies on prompt-based generation, and the adopted IPA strategy may not be optimal; more advanced and structured prompt formulation techniques could further improve personalized generation performance. Nevertheless, prompt engineering, which is crucial for performance
of LLMs, is not the central objective of this study. Second, the LaMP-1 (Personalized Citation Identification) and LaMP-5 (Personalized Scholarly Title Generation) tasks provide only paper abstracts rather than full-text content, which may limit the contextual information available to LLMs when selecting citations or generating titles, even though abstracts offer useful sequence constraints. Third, our evaluation is restricted to English, text-only datasets, leaving the effectiveness of ClusterRAG in multilingual and multimodal settings unexplored. Finally, ClusterRAG’s end performance depends on the underlying language model, which may introduce additional limitations inherited from the backend LLM.

Future work will focus on developing more advanced prompt generation strategies and extending ClusterRAG to support multilingual and multimodal personalized retrieval-augmented generation (RAG) settings. We also plan to broaden the scope of generation by incorporating a wider range of generative model families beyond the current Flan and Qwen variants. Another promising direction involves integrating feedback from the generative model into the retrieval process to further enhance system performance. In addition, future research on user similarity computation will explore more efficient and adaptive clustering methodologies, including periodic batch re-clustering, incremental clustering approaches, and online embedding updates with dynamic cluster reassignment. We will also investigate strategies for maintaining cluster centroids and assigning new users without requiring full recomputation.

\section*{Ethical Considerations}\label{sec:ethical_considerations}
ClusterRAG leverages user interaction histories and collaborative signals, which raises considerations related to privacy, consent, and potential bias. Although the framework operates on anonymized user profiles and does not require access to explicit personal identifiers, improper handling of user-generated data could still risk unintended information leakage. Moreover, collaborative filtering may amplify existing biases if certain user groups or preferences are overrepresented in the data, potentially affecting fairness in personalized outputs.

To mitigate these risks, ClusterRAG can be deployed with standard data governance practices, including data anonymization, access control, on-device embedding computation, secure aggregation, and bias-aware evaluation. Importantly, ClusterRAG is a model-agnostic retrieval framework rather than a user profiling system, and it does not infer sensitive attributes (e.g., user demographics) beyond observed interactions. Additionally, ClusterRAG operates on embedding-level user representations rather than raw textual logs during clustering and similarity ranking, and constructs user profiles based solely on historical interactions rather than demographic information, helping limit ethical exposure while enabling effective personalization.

\section*{Acknowledgments}
This work is supported by the National Science Foundation (NSF) under Award number OIA-1946391, Data Analytics that are Robust and Trusted (DART).

\bibliography{custom}

\appendix

\section{Prompts Used in ClusterRAG}\label{sec:prompts_appendix}
This subsection presents the prompt templates employed during generation for the remaining LaMP tasks. As already stated, each prompt contains an instruction, input (query), and profile. We provide the templates below. In the templates,  \{\textit{Movie description}\} and \{\textit{Movie tags}\}, \{\textit{Review}\}, \{\textit{Article}\}, \{\textit{Paper abstract}\}, and \{\textit{Tweet}\} represent user input for the corresponding LaMP tasks, while the rest of the italicized text represent user profile entries. The remaining text is the instruction guiding an LLM to generate the intended output.

\begin{tcolorbox}
\textbf{LaMP-2 (Personalized Movie Tagging) Prompt Template:} %\medskip 
Given the user previous movie tag pairs:
The tag for the movie description: <\textit{Movie\_1\_Description}> is <Tag\_1>, the tag for the movie description: <\textit{Movie\_2\_Description}> is <\textit{Tag\_2}>, $\dots$, the tag for the movie description: <\textit{Movie\_M\_Description}> is <\textit{Tag\_M}>, which tag does the  movie description: \{\textit{Movie description}\}  relate to among the following tags? Just answer with the tag name without further explanation. Movie tags:
\{\textit{Movie tags}\}
\end{tcolorbox}

\begin{tcolorbox}
\textbf{LaMP-3 (Personalized Product Rating) Prompt Template:}
%\medskip 
Given the user previous review-score pairs:  
<\textit{Score\_1}> is the score for <\textit{Review\_1 Text}>,  
<\textit{Score\_2}> is the score for <\textit{Review\_2 Text}>, \dots, <\textit{Score\_M}> is the score for <\textit{Review\_M Text}>. What is the score of the following review on a scale of 1 to 5? Just answer with 1, 2, 3, 4, or 5 without further explanation. Review: \{\textit{Review}\}
\end{tcolorbox}

\begin{tcolorbox}
\textbf{LaMP-4 (Personalized News Headline Generation)  Prompt Template:}
%\medskip
Given the user’s previous article-headline pairs:  
<\textit{Headline\_1}> is the title for <\textit{Article\_1\_Text}>,  
<\textit{Headline\_2}> is the title for <\textit{Article\_2\_Text}>, \dots, <\textit{Headline\_M}> is the title for <\textit{Article\_M\_Text}>. Generate a headline for the following article. Article: 
\{\textit{Article}\}
\end{tcolorbox}

\begin{tcolorbox}
\textbf{LaMP-5 (Personalized Scholarly Title Generation) Prompt Template:}
%\medskip
Given the user’s previous abstract-title pairs:  
<\textit{Title\_1}> is a title for <\textit{Abstract\_1\_Text}>,  
<\textit{Title\_2}> is a title for <\textit{Abstract\_2\_Text}>, \dots, <\textit{Title\_M}> is a title for <\textit{Abstract\_M\_Text}>. Generate a title for the following abstract of a paper. Abstract:\{\textit{Paper abstract}\}  
\end{tcolorbox}

\begin{tcolorbox}
\textbf{LaMP-7 (Personalized Tweet Paraphrasing) Prompt Template:}
%\medskip
Given the user’s previous tweets:  
<\textit{Tweet\_1}>, <\textit{Tweet\_2}>, \dots, <\textit{Tweet\_M}>. Paraphrase the following tweet without any explanation before or after it following the user's tweeting patterns. Tweet: \{\textit{Tweet}\}
\end{tcolorbox}

\section{Detailed Dataset Statistics and Licensing Information}\label{sec:detailed_lamps_statistics}

Table~\ref{tab:lamp_stats_appendix} below provides detailed statistics of the LaMP benchmark based on the time-based split, as stated in Section~\ref{exp_setup}. 
We additionally summarize the licensing information and terms of use for each LaMP task considered in this study as follows:
\begin{enumerate}
    \item Personalized Citation Identification (LaMP-1): CC BYNC-SA 4.0.
    \item Personalized Movie Tagging (LaMP-2): Educational or academic research, NON COMMERCIAL USE.
    \item Personalized Product Rating (LaMP-3): CC BY-NC-SA 4.0.
    \item Personalized news Headline Generation (LaMP-4): CC BY-NC-SA 4.0.
    \item Personalized Scholarly Title Generation (LaMP-5): CC BY-NC-SA 4.0.
    \item Personalized Tweet Paraphrasing (LaMP-7): CC BYNC-SA 4.0.
\end{enumerate}

\begin{table*}[t]
\centering
\resizebox{\textwidth}{!}{
\begin{tabular}{lcccccccc}
\toprule
\textbf{Task} & \textbf{\#users} & \textbf{\#train} & \textbf{\#dev} & \textbf{\#test} & \textbf{Input Length} & \textbf{Output Length} & \textbf{\#Profile Size} & \textbf{\#classes} \\
\midrule
LaMP-1 & 6542  & 6542  & 1500 & 1500 & 51.43 $\pm$ 5.70   & --               & 84.15 $\pm$ 47.54   & 2  \\
LaMP-2 & 929   & 5073  & 1410 & 1557 & 92.39 $\pm$ 21.95  & --               & 86.76 $\pm$ 189.52  & 15 \\
LaMP-3 & 20000 & 20000 & 2500 & 2500 & 128.18 $\pm$ 146.25 & --              & 185.40 $\pm$ 129.30 & 5  \\
LaMP-4 & 1643  & 12500 & 1500 & 1800 & 29.97 $\pm$ 12.09  & 10.07 $\pm$ 3.10 & 204.59 $\pm$ 250.75 & -- \\
LaMP-5 & 14682 & 14682 & 1500 & 1500 & 162.34 $\pm$ 65.63 & 9.71 $\pm$ 3.21  & 87.88 $\pm$ 53.63   & -- \\
LaMP-7 & 13437 & 13437 & 1498 & 1500 & 29.72 $\pm$ 7.01  & 16.96 $\pm$ 5.67 & 15.71 $\pm$ 14.86   & -- \\
\bottomrule
\end{tabular}
}
\caption{Detailed statistics of the LaMP benchmark with time-based data split.}
\label{tab:lamp_stats_appendix}
\end{table*}

\section{Task Descriptions}\label{app_sec:appendix1}

The LaMP benchmark contains English and text-only data. The documents used in each LaMP task do not contain personally identifiable information that could otherwise compromise privacy issues. Below, we provide detailed descriptions of each LaMP task included in our evaluation, outlining the task objectives, input-output formulations, and the specific aspects of personalization each task is designed to assess. 

\paragraph{LaMP-1: Personalized Citation Identification.}
This task frames citation recommendation as a binary
classification task and assesses the ability of a language model to identify user preferences for citations. Specifically, if the user $u$ writes a paper $x$, a language model must determine which of two given candidate papers ($a$ or $b$) $u$ will cite in $x$.  The profile of each user encompasses all the papers they have authored. Only the title and abstract of each paper are retained in the user’s profile; it uses scientific papers.

\paragraph{LaMP-2: Personalized Movie Tagging.}
This task recasts movie tagging as a multi-class classification task. Given a movie description
$x$ and a user’s historical movie-tag pairs, a
language model must predict one of 15 tags for $x$. The movie tags are: sci-fi, based on a book, comedy, action, twist ending, dystopia, dark comedy, classic, psychology, fantasy, romance, thought-provoking, social commentary, violence, and true story.

\paragraph{LaMP-3: Personalized Product Rating.}
LaMP-3 is also framed as a multi-class classification task. In particular, given the user $u$’s historical review and rating pairs of products and an input review $x$, the model must predict an integer rating (from 1 to 5) of the review.

\paragraph{LaMP-4: Personalized News Headline Generation.} 
This is a generative task that evaluates the ability of an LLM to capture the stylistic patterns of an author $u$ by requiring it to generate a headline for an input news article, $x$,
given a user profile of the authors’ historical article-title pairs.

\paragraph{LaMP-5: Personalized Scholarly Title Generation.}
LaMP-5 is another generative task that requires a generative model to generate titles for an input article $x$, given a user profile of historical article-title pairs for an author. It is similar to LaMP-4, but it uses different corpus domain, scientific papers (similar to LaMP-1). 

\paragraph{LaMP-7: Personalized Tweet Paraphrasing.}

This is framed as a generative personalized tweet paraphrasing task, which requires an LLM  to generate a tweet in the style of a user $u$ given an input tweet $x$, and a user profile of historical tweets by the
user.

\section{Hyperparameter Tuning Grid}\label{sec:appendix2}
Table~\ref{tab:hyperparam_tune_grid} summarizes the hyperparameter search space explored during model selection, detailing the ranges and candidate values used to tune ClusterRAG across optimization, training, and decoding configurations. As described in Section~\ref{exp_setup}, the optimal hyperparameters were identified via grid search with early stopping applied on LaMP-1 and LaMP-7, representing classification and generative tasks, respectively. 

\begin{table}[t]  
\centering
\renewcommand{\arraystretch}{1.15}
\footnotesize  
\begin{tabular}{ll}  
\toprule  
\textbf{Hyperparameter} &  \textbf{Tested values}   \\  
\midrule 

Learning rate &	 5 $\times10^{-5}$,  $3 \times 10^{-3}$, $10^{-3}$,  $10^{-4}$ \\
Weight decay &	5 $\times$ $10^{-6}$, 10$^{-4}$, $10^{-3}$ \\
Warm-up ratio & 0.05 to 0.10 \\
Batch size &	8, 16, 32, 64\\
Epochs &	10, 20 30, 50, 70, 100\\
Max seq length &  64, 128, 256, 512 \\
Beam size & 1 $-$ 6  \\ 
$\gamma$ &	 0.1 $-$ 0.9\\
$k$   & 1 $-$ 5 \\
$m$   & 1 $-$ 12\\
$L_{\max}$ & 64, 128, 256, 512, 1024 \\
$|\bar{y}|$ & 32, 64, 128, 256, 512 \\
\bottomrule
\end{tabular}%  
  
 \caption{Hyperparameter tuning grid used for training and optimizing ClusterRAG.} 
\label{tab:hyperparam_tune_grid} 
\end{table}

\section{Impact of Similar User Size and Profile Size} \label{sec:appendx_impact_k_m}

This section investigates the sensitivity of ClusterRAG to two key design parameters that govern collaborative context construction: the number of similar users ($k$) incorporated during retrieval and the number of profile documents ($m$) selected per user. By systematically varying these parameters, we analyze how the breadth of collaborative signals and the depth of user profile information affect model performance across different LaMP tasks.

\begin{table*}[t]
\centering
\setlength{\tabcolsep}{3.8pt}
\renewcommand{\arraystretch}{1.15}
\footnotesize
\begin{tabular}{lcccccccccccc}
\toprule
\textbf{$k$ value} 
& \multicolumn{2}{c}{\textbf{LaMP-1}} 
& \multicolumn{2}{c}{\textbf{LaMP-2}} 
& \multicolumn{2}{c}{\textbf{LaMP-3}} 
& \multicolumn{2}{c}{\textbf{LaMP-4}} 
& \multicolumn{2}{c}{\textbf{LaMP-5}} 
& \multicolumn{2}{c}{\textbf{LaMP-7}} \\
\cmidrule(lr){2-3}\cmidrule(lr){4-5}\cmidrule(lr){6-7}
\cmidrule(lr){8-9}\cmidrule(lr){10-11}\cmidrule(lr){12-13}

& Acc.$\uparrow$ & F1$\uparrow$
& Acc.$\uparrow$ & F1$\uparrow$
& MAE$\downarrow$ & RMSE$\downarrow$
& R-1$\uparrow$ & R-L$\uparrow$
& R-1$\uparrow$ & R-L$\uparrow$
& R-1$\uparrow$ & R-L$\uparrow$ \\
\midrule
1 & 0.690 & 0.690 & 0.661 & 0.620 & 0.270 & 0.594 & 0.190 & 0.176 & 0.490 & 0.440  & 0.521 & \textbf{0.470} \\

2 &0.692 & 0.697 & 0.665 & 0.631 & \textbf{0.269} & \textbf{0.582} & 0.193 & 0.179 & 0.496 & 0.444  & 0.524 & 0.469   \\

3 &\textbf{0.700} & \textbf{0.700} & \textbf{0.668} & \textbf{0.642} & 0.270 & 0.595 &\textbf{ 0.196} & \textbf{0.180} & 0.495 & \textbf{0.445}  & \textbf{0.528} & 0.468 \\

4  & 0.690 & 0.688 & 0.660 & 0.621 & 0.272 & 0.595 & 0.192 & 0.178 & \textbf{0.497} & 0.441  & 0.523 & 0.469  \\

5  &0.689 & 0.690 & 0.658 & 0.619 & 0.273 & 0.596 & 0.191 & 0.177 & 0.492 & 0.438  & 0.521 & 0.467  \\

\bottomrule
\end{tabular}
\caption{Effect of the number of similar users ($k$) on ClusterRAG performance in hybrid mode across LaMP tasks.}
\label{tab:impact_k}
\end{table*}

\begin{table*}[t]
\centering
\setlength{\tabcolsep}{3.8pt}
\renewcommand{\arraystretch}{1.15}
\footnotesize
\begin{tabular}{lcccccccccccc}
\toprule
\textbf{$m$ value} 
& \multicolumn{2}{c}{\textbf{LaMP-1}} 
& \multicolumn{2}{c}{\textbf{LaMP-2}} 
& \multicolumn{2}{c}{\textbf{LaMP-3}} 
& \multicolumn{2}{c}{\textbf{LaMP-4}} 
& \multicolumn{2}{c}{\textbf{LaMP-5}} 
& \multicolumn{2}{c}{\textbf{LaMP-7}} \\
\cmidrule(lr){2-3}\cmidrule(lr){4-5}\cmidrule(lr){6-7}
\cmidrule(lr){8-9}\cmidrule(lr){10-11}\cmidrule(lr){12-13}

& Acc.$\uparrow$ & F1$\uparrow$
& Acc.$\uparrow$ & F1$\uparrow$
& MAE$\downarrow$ & RMSE$\downarrow$
& R-1$\uparrow$ & R-L$\uparrow$
& R-1$\uparrow$ & R-L$\uparrow$
& R-1$\uparrow$ & R-L$\uparrow$ \\
\midrule
1 &0.674  &0.673  &0.648  &0.612  &0.289 &0.608  &0.182  &0.166  &0.480  &0.431  &0.515  &0.464  \\

2 & 0.690 & 0.690 & 0.661 & 0.620 & 0.270 & 0.594 & 0.190 & 0.176 & 0.490 & 0.440  & 0.521 & 0.470 \\

3  &0.691 & 0.691 & 0.665 & 0.632 & 0.269  & 0.591 & 0.191 & 0.178 & 0.501 & 0.465 & 0.528 &0.474 \\

4 &0.693  &0.692  &0.669  &0.639  &0.268	& 0.590  &0.196  &0.179  &0.510  &0.471  &0.530  &0.473\\

5  &0.695  &0.695  &\textbf{0.675}  &\textbf{0.650}  &0.267  &0.586  &0.198  &0.180 &0.512  &0.473  &0.533  &0.475  \\

6  &\textbf{0.707}  &\textbf{0.707}  &0.673  &0.649  &0.262  &0.584  &\textbf{0.200}  &\textbf{0.182} &\textbf{0.514}  &\textbf{0.479}  &0.534  &0.476  \\

7  &0.703  &0.700  &0.672  &0.647  &\textbf{0.260}  &\textbf{0.581}  &0.199  &0.180 &0.511  &0.474  &\textbf{0.538}  &\textbf{0.481}  \\

8  &0.704  & 0.701  &0.670  &0.645  &0.263  &0.583  &0.197  &0.178 &0.509  &0.472  &0.536  &0.478  \\

9  &0.707  &0.701  &0.671  &0.643  &0.263  &0.585  &0.196  &0.177 &0.511  &0.474  &0.534  &0.476  \\

10  &0.706  &0.705  &0.671  &0.644  &0.261  &0.582  &0.191  &0.174 &0.508  &0.471  &0.533  &0.476  \\

11  &0.701  &0.700  &0.669  &0.641  &0.263  &0.584  &0.191  &0.173 &0.504  &0.469  &0.534  &0.475  \\

12  &0.700  &0.700  &0.670  &0.664  &0.264  &	0.584  &0.192  &0.174 &0.504  &0.470  &0.532  &0.473  \\

\bottomrule
\end{tabular}
\caption{Effect of the number of retrieved profile documents ($m$) on ClusterRAG performance in hybrid mode across LaMP tasks.}
\label{tab:impact_m}
\end{table*}

\paragraph{ Impact of the Number of Similar Users ($k$).} 
Table~\ref{tab:impact_k} shows that incorporating a small number of similar users consistently improves ClusterRAG performance across all LaMP tasks. Performance generally increases as $k$ grows from 1 to 3, where most metrics achieve their peak or near-peak values, indicating that a limited set of highly similar users provides the most informative collaborative signals. Beyond $k=3$, gains saturate or slightly decline, suggesting that adding more users introduces weaker or noisier preferences that dilute personalization benefits. This trend highlights the importance of selectively leveraging collaborative information rather than aggregating large numbers of loosely related users.

\paragraph{Impact of the Number of Retrieved Profile Documents ($m$).} As shown in Table~\ref{tab:impact_m}, increasing the number of retrieved profile documents leads to steady performance improvements up to a moderate range ($m\approx6$–$7$), after which the gains plateau. Larger profile sizes consistently reduce prediction error (MAE/RMSE) and improve generation quality (ROUGE scores), reflecting richer contextual grounding. At the same time, marginal benefits diminish for larger $m$, indicating that excessively long profiles provide limited additional signal while increasing prompt complexity. Overall, these results suggest that ClusterRAG is robust to the choice of $m$ and performs best when balancing sufficient contextual coverage with prompt efficiency.

\section{Cluster Cohesion for Collaborative User Retrieval}\label{sec_app:cluster_cohesion}

To evaluate whether ClusterRAG forms meaningful user clusters for collaborative filtering, we measure the Silhouette score~\citep{ROUSSEEUW198753} of user clusters produced by HDBSCAN and $k$-means and  report results in Table~\ref{tab:cluster_quality}. The Silhouette score is an internal clustering metric that jointly captures intra-cluster cohesion and inter-cluster separation, with values ranging from $-1$ to $1$, where higher scores indicate better-defined clusters. 

As shown in Table~\ref{tab:cluster_quality}, ClusterRAG consistently achieves higher Silhouette scores with HDBSCAN than with $k$-means, whose scores remain below $0.5$ across all tasks. In high-dimensional embedding spaces, moderately positive Silhouette scores are common and still reflect meaningful structure. Therefore, the scores above $0.5$ obtained with HDBSCAN indicate that ClusterRAG learns cohesive and well-separated user clusters, enabling the retrieval of more similar users for collaborative filtering in personalized RAG. These findings further corroborate the superior performance of ClusterRAG when combined with HDBSCAN.

\begin{table}[t]  
\centering
\renewcommand{\arraystretch}{1.15}
\footnotesize  
\begin{tabular}{lcc}  
\toprule  
\textbf{Task} &  \textbf{HDBSCAN Score}  & \textbf{$k$-means  Score} \\  
\midrule 

LaMP-1 &0.601  &0.389 \\
LaMP-2 &0.535  & 0.326 \\
LaMP-3 &0.551  & 0.328 \\
LaMP-4 &0.570  &0.274\\
LaMP-5 &0.562  &0.347\\
LaMP-7 & 0.537  &0.323\\

\bottomrule
\end{tabular}%  
  
 \caption{Silhouette scores of user clusters produced by ClusterRAG using HDBSCAN and $k$-means on the LaMP benchmark.}
\label{tab:cluster_quality} 
\end{table} 

\section{Case Study}
\label{sec:case_study}

\begin{table*}[t]
\centering
\renewcommand{\arraystretch}{1.25}
\setlength{\tabcolsep}{6pt}
\begin{tabular}{p{0.95\textwidth}}
\midrule
\textbf{User Query:} \\
\textbf{Movie description}: When the Davison family comes under attack during their wedding anniversary getaway, the gang of mysterious killers soon learns that one of their victims harbors a secret talent for fighting back. \\
\midrule
\textbf{Gold Output:} Violence \\
\midrule
\textbf{Target User Profile:} \\
     (1) \textbf{Movie tag}: ``twist ending'', 
    \textbf{Movie description}: ``Soon after his insufferably arrogant father wins the Nobel Prize for chemistry, Barkley Michaelson is kidnapped by Thaddeus James, a young genius who claims to be Barkley's illegitimate half-brother. Motivated not so much by money as revenge, Thaddeus tries to convince Barkley to help him carry out a multimillion-dollar extortion plot against their patriarch.''\\

    (2)\textit{ \textbf{Movie tag}: ``action'', 
    \textbf{Movie description}: ``The Bride unwaveringly continues on her roaring rampage of revenge against the band of assassins who had tried to kill her and her unborn child. She visits each of her former associates one-by-one, checking off the victims on her Death List Five until there's nothing left to do \dots but kill Bill.''} \\

    (3) \textbf{Movie tag}: ``action'', 
    \textbf{Movie description}: ``NYPD cop John McClane's plan to reconcile with his estranged wife is thrown for a serious loop when, minutes after he arrives at her office, the entire building is overtaken by a group of terrorists. With little help from the LAPD, wisecracking McClane sets out to single-handedly rescue the hostages and bring the bad guys down.'' \\
\midrule
\textbf{Similar User Profile:} \\
    (1) \textbf{Movie tag}: ``true story'', 
    \textbf{Movie description}: ``The mostly true story of the legendary `worst director of all time', who, with the help of his strange friends, filmed countless B-movies without ever becoming famous or successful.''\\

    (2) \textbf{Movie tag}: ``action'', 
    \textbf{Movie description}: ``Liu Jian, an elite Chinese police officer, comes to Paris to arrest a Chinese drug lord. When Jian is betrayed by a French officer and framed for murder, he must go into hiding and find new allies.''\\

     (3) \textit{\textbf{Movie tag}: ``violence'', 
    \textbf{Movie description}: ``An elderly ex-serviceman and widower looks to avenge his best friend's murder by doling out his own form of justice.''}\\
\midrule
\textbf{Top Ranked Documents:} 
Doc \# (3) from similar user then doc \# (2) from current user.\\
\midrule

\textbf{Personalized Prompt:}\\

Given the user previous movie tag pairs:
The tag for the movie description: ``\textit{An elderly ex-serviceman and widower looks to avenge his best friend's murder by doling out his own form of justice}'' is ``\textit{violence}'', and the tag for the movie description: ``\textit{The Bride unwaveringly continues on her roaring rampage of revenge against the band of assassins who had tried to kill her and her unborn child. She visits each of her former associates one-by-one, checking off the victims on her Death List Five until there's nothing left to do \dots but kill Bill.}'' is ``\textit{action}'', which tag does the  movie description: \textit{``When the Davison family comes under attack during their wedding anniversary getaway, the gang of mysterious killers soon learns that one of their victims harbors a secret talent for fighting back.}''  relate to among the following tags? Just answer with the tag name without further explanation. Movie tags:
``\textit{sci-fi, based on a book, comedy, action, twist ending, dystopia, dark comedy, classic, psychology, fantasy, romance, thought-provoking, social commentary, violence, and true story}''\\

\midrule
\textbf{Generated Output}: Violence \\
\midrule
\end{tabular}
\caption{A  case study illustrating how ClusterRAG leverages user profile and similar-user information for personalized RAG.}
\label{tab:case_study}
\end{table*}

We randomly sample a case from \textbf{LaMP\_2} \textbf{(Personalized Movie Tagging)} in Table~\ref{tab:case_study} to illustrate the effectiveness of ClusterRAG in leveraging both target and similar user profiles for personalized generation. In this task, the \textit{User Query} corresponds to a movie description, and the objective is to generate an appropriate movie tag based on this description and the user’s historical tagging behavior. Due to space constraints, we include three profile documents per user. The target user’s historical tags are \textit{twist ending} and \textit{action}, while the similar user’s historical tags include \textit{true story}, \textit{action}, and \textit{violence}. The gold label for the given query is \textit{violence}.

When we use a non-personalized prompt or  the target user profile only, ClusterRAG  incorrectly predicts the movie tag as \textit{action}, driven by its higher frequency in the current user’s history. However, when similar-user information is incorporated via hybrid profile retrieval, the model correctly predicts \textit{violence}, as the movie tag. This occurs because the top-ranked retrieved profile, originating from a similar user, emphasizes personal and retaliatory violence that closely aligns with the query, whereas the lower-ranked \textit{action}-tagged profile reflects more stylized narratives less relevant to the description.

\section{AI Assistance Usage}
In this work,  ChatGPT has been used solely as a writing assistant. Specifically, draft passages were provided to the tool for paraphrasing and language refinement, after which we manually reviewed, edited, and finalized the text.

\end{document}